# A solution to the proton and deuteron size puzzle and the anomalous magnetic moment of the muon

Jean Paul Mbelek[1]

1. University of Douala, Douala, Cameroon

Correspondence to: Jean Paul Mbelek[1] (Email: mbelek.jp@wanadoo.fr).

Abstract : We show that a proposal which involves an effective radius derived from an effective potential that includes a linear extra-potential may solve both the proton and deuteron size puzzle. Moreover, this solution preserves the e-μ universality and helps to solve the anomalous magnetic moment of the muon too. Still, it is consistent with the rms magnetic radius of the proton and deuteron, the rms charge radius of both the ordinary and muonic C12 as well as the kaonic and pionic X-rays spectroscopy.

## I. Introduction

A few teams [1, 2, 3, 4] have made the most precise measurements of the charge radius of the proton to date. The experiments based on electrons beams yield the proton radius either from scattering experiments or by measuring the energy levels of ordinary hydrogen atoms. The experiments based on muons beams as yet have deduced the proton charge radius only by measuring the hyperfine (HF) energy levels of muonic hydrogen atoms. Now, the experiments based on electron scattering [3, 4] or the electron energy levels agree with the quantum electrodynamics (QED) expectation $\Delta E^{QED}_{2P3/2(F=2)-2S1/2(F=1)}$ = 205.9835 meV whereas the one based on the Lamb shift in muonic hydrogen yields $\Delta E^{exp}_{2P3/2(F=2)-2S1/2(F=1)}$ = 206.2949(32) meV [1] and $\Delta E^{exp}_{2P3/2(F=2)-2S1/2(F=1)}$ = 206.2927(27) meV [2]. The discrepancy amounts to (0.314 ± 0.007) meV between the QED prediction and experiments. The root-mean square charge radius of the proton, $r_p = (<r_p^2>)^{1/2}$, is derived by solving the QED equation $\Delta E = 209.9779(49) - 5.2262\, r_p^2 + 0.0347\, r_p^3$, where the energy difference between the $2S_{1/2}^{F=1}$ and $2P_{3/2}^{F=2}$ states, ΔE, and $r_p$ are respectively expressed in millielectronvolt (meV) and femtometre (fm). The last measurements of the muonic hydrogen energy levels first yielded $r_p$ = (0.84184±0.00067) fm, now confirmed to $r_p$ = (0.84087±0.00039) fm, which is 5$\sigma$ to 7$\sigma$ lower than the QED prediction as expressed by the CODATA values $r_p$ = (0.8768±0.0069) fm (2008 CODATA value) [4] or $r_p$ = (0.8751±0.0061) fm (2010 CODATA value) [5] or $r_p$ = (0.879±0.011) fm (2014 CODATA value) [6]. The authors conclude from their result that either the Rydberg constant, $R_\infty$, has to be shifted by −110 kHz/*c* (4.9 standard deviations), or the calculations of the QED effects in atomic hydrogen or muonic hydrogen atoms are insufficient [1]. However, some other authors claimed that such an explanation is disfavored because of the true difference between electron-based and muon-based measurements [7]. It seems rather that the results of the experiments depend on the flavor or the generation of the charged lepton that is used to probe the proton differing by more than five times the uncertainty in either measurement thereby suggesting new physics [8,9,10,11,12] (see also [13], for a review). Now, no evidence of new physics has been reported at the LHC and many constraints from low energy data disfavor new spin-0, spin-1 and spin-2 particles as an explanation [14]. Perhaps, experiments are just suggesting that a correction involving either the mass or the weak hypercharge or isospin instead of the flavor of the test lepton is needed in order to make the measurements of the proton charge radius using electrons and those using muons as probes be consistent with one another. Such an effect might be pointing out for a non trivial connection between QED and QCD through the ratio $(m/M)^2$ [15]. The latter is also expected as implied by new physics in the study of the the anomalous magnetic moment [16] (see also [17], Section 3.2.3, p. 169 or arXiv:hep-ph/0002158, Section XXXIIIC, p.177). For instance, the virtual photon may be in competition with another virtual massless neutral vector boson, namely the gluon, while the nucleons interact with a charged lepton. We put forward the Feynman diagrams of FIG.1 and FIG. 2 to emphasize such a mixing possibility[1]. It means that a non trivial hadronic vacuum polarization correction might be taken into account in order to solve the proton and the deuteron radius puzzle as well as the anomalous magnetic moment of the muon. The general idea that we emphasize throughout in this paper is to show how a replacement of the kind $1/r^n \rightarrow 1/r_{eff}^n = f(\Lambda r/ƛMc^2)/r^n$ may help to solve the light nuclei radii puzzle for experiments in which the test particle, namely a charged lepton, probes the vicinity of a nucleus down to a radius r of order or less than ƛ, where ƛ = $(\hbar/mc)$ is the reduced Compton wavelength of the lepton and m is its mass. The function f is endowed with a local minimum point at r = 0 ($f'(0) = 0$ and $f''(0) > 0$) and is such that $f(0) = 1$. For r « ƛ$(Mc^2/\Lambda)$, it follows $f(\Lambda r/ƛMc^2) \approx 1 + \zeta^2 [(m/M)(\Lambda r/\hbar c)]^2$, where we have set $\zeta^2 = ½ f''(0)$. These considerations give rise to a linear extra-potential, $V_L(r) = \hbar c Z\alpha \zeta^2 (m/M)^2 (\Lambda/\hbar c)^2 r$ in addition to the usual Coulomb potential, $V_C(r) = -(\hbar c Z\alpha/r)$, where $\alpha = (e^2/4\pi\varepsilon_0\hbar c)$ is the fine structure constant. In this paper, we do not address the set of new rules that could lead to the corrected relation $\Delta E = 209.9779(49) - 5.2262\, r_{peff}^2 + 0.0347\, r_{peff}^3$ from the inclusion of the extra-potential, $V_L(r)$, by starting from the Hamiltonian in a semiclassical approach. For the time being, we just consider that the effective proton (or deuteron) charge radius, $r_{peff}$, should replace the radius, $r_p$, of the Minkowski metric in any result where the radius is involved. Besides, let us emphasize that by making the replacement $1/r^n \rightarrow 1/r_{eff}^n$, we should consider a spacetime dependent metric $g_{\mu\nu}$ instead of the usual Minkowski flat metric, $\eta_{\mu\nu}$. Now, the $g_{\mu\nu}$'s too satisfy the vacuum Einstein equations. So, $g_{\mu\nu}$ merely defines a flat metric conformal to the Minkowski metric, namely $g_{\mu\nu} = \Omega^2 \eta_{\mu\nu}$. Thus, the geometry of spacetime is flat within a nucleus and in its outskirt as well up to a Weyl rescaling. Indeed, the density of a hydrogen nucleus is of the order of $6\times10^{17}$ kg $m^{-3}$ whereas the density of a hydrogen atom is of the order of $2.7\times10^3$ kg $m^{-3}$. We consider that the conformal transformation $g_{\mu\nu} = \Omega^2 \eta_{\mu\nu}$ is supported by such a difference of fourteen order of magnitude. As a consequence, the above considerations imply a new correction to the Coulomb potential. Following the views already pointed out by some authors [17,18], this correction is mixed as $\alpha_s (Z\alpha) (m/M)^2$, where $\alpha_s$, $\alpha$, Z, m and M denote respectively the strong coupling constant, the fine structure constant, the atomic number, the mass of the charged lepton and the mass of the nucleus. As one can see a product of factors of the kind $\alpha_s^x (Z\alpha)^y (m/M)^{x+y}$ can be looked at as a product of effective coupling constants, namely $[\alpha_s (m/M)]^x$ and $[(Z\alpha) (m/M)]^y$. Besides, although leptons are not by themselves sensitive to the strong interaction, the latter may perturb the QED (or the EW) interaction of leptons (see FIG. 1 or FIG. 2). This is the case especially for the ground states of the muonic hydrogen and the muonic deuterium where the muon may probe the inner part of the nucleus with a non-zero probability. Moreover, the muon can interact with each quark of a nucleon not only with the nucleus as a whole. All these remarks lead us to consider missing corrections involving either QED and QCD or EW and QCD mixing terms. Finally, the same idea may help to solve the problem of the anomalous magnetic moment of the muon too. The observed discrepancy amounts to $\delta a_\mu = a_\mu^{exp} - a_\mu^{SM} = (302 \pm 87) \times 10^{-11}$ [19,20,21,22], whereas the total SM prediction is $a_\mu^{SM} = (11659180.2 \pm 5.3) \times 10^{-10}$ [19,20,22], with the QED contribution $a_\mu^{QED} = (116584718.09\pm0.16)\times10^{-11}$ [14,15,20,22,23], the hadronic contribution $a_\mu^{had} = (6921\pm56)\times10^{-11}$ [19,22,24,25] and the electroweak corrections $a_\mu^{EW} = (154\pm2)\times10^{-11}$ [21,22,26]. The plan of the paper is as follows: In section II the derivation of the linear extra-potential is reviewed. In section III the proton charge radius from muonic hydrogen and the deuteron charge radius from muonic deuterium are computed by using the effective radius implied by the linear extra-potential. In section IV the magnetic radius of ordinary and muonic hydrogen are computed by using the effective radius implied by the linear extra-potential. In section V the anomalous magnetic moment of the muon is estimated still based on the linear extra-potential. Finally, we conclude in section VI.

[1] Feynman diagrams analogous to those of FIG. 1 and 2 and with only QED coupling could be considered but with an amplitude at least ten times weaker since $\alpha$ « $\alpha_s$. Moreover, because in that case $g_{\mu\nu} = \eta_{\mu\nu}$ exactly at all points, this would give rise to a potential term that decreases as $1/r^2$ and can by no means be a linear term.

II. Derivation of the linear extra-potential

As one knows, the static potential, V, is the inverse Fourier transform of the scattering amplitude, $T_{fi}$ [27]. As usual, $(\Upsilon^{\mu} q_{\mu} + m_q)/(\mathbf{q}^2 - m_q^2)$, $- i\, g_{\mu\nu}\, \delta^{ab}/k'^2$ and $- i\, g_{\mu\nu}/k^2$ are respectively the quark, the gluon and the photon propagators in the Feynman gauge. Let us set $\hbar = c = 1$ and perform the conformal transformation $\eta_{\mu\nu} \rightarrow g_{\mu\nu} = \Omega(\mathbf{r})^2\, \eta_{\mu\nu}$ in analogy with the conformal correspondence between the Einstein frame and the Jordan frame. Thus, $g_{\mu\nu}$ defines a flat metric conformal to the Minkowski metric. Besides, let us introduce the "mixing" coupling constants $\alpha^* = \alpha\, \zeta\, (m/M)$ and $\alpha_s^* = \alpha_s\, \zeta\, (m/M)$, where the ratio of the masses of the nucleus-charged lepton two-body system, m/M, accounts for the recoil correction. In the same manner, the quantity $\zeta$ enters in the linear part of the potential to account for the isotopic effect. Thus, the scattering amplitude is given by $T_{fi} = T_{fi}^{(0)} + T_{fi}^{(1)}$, where $T_{fi}^{(0)} = (1/2\pi)^6\, Z\alpha/\mathbf{k}^2$ is the QED contribution and

$$T_{fi}^{(1)} = \tfrac{3}{4} \int \Omega(\mathbf{r})^4\, (1/2\pi)^6\, (Z\alpha^*/\mathbf{k}^2)\, (\alpha_s^*/\mathbf{k}'^2)\, [(\Upsilon^{\mu} q_{\mu} + m_q)/(\mathbf{q}^2 - m_q^2)]\, [(\Upsilon^{\mu} q'_{\mu} + m_q)/(\mathbf{q}'^2 - m_q^2)]\, d^3\mathbf{r}$$

is the mixing term described in FIG. 1 that involves both the photon, $\Upsilon$, and the gluon, G, propagators altogether.

Then it follows $V(r) = V_C(r) + V_L(r)$, where

$$V_C(r) = - Z\alpha\, (1/2\pi)^3 \int e^{-i\mathbf{k}.\mathbf{r}}\, d^3\mathbf{k}/\mathbf{k}^2 \quad (1)$$

and

$$V_L(r) = - \tfrac{3}{4}\, Z\alpha^*\, \alpha_s^*\, (1/2\pi)^3 \int\int\int \Omega^4\, e^{-i\mathbf{k}.\mathbf{r}}\, [(\Upsilon^{\mu} q_{\mu} + m_q)(\Upsilon^{\mu} q'_{\mu} + m_q)/(\mathbf{q}^2 - m_q^2)(\mathbf{q}'^2 - m_q^2)]\, (d^3\mathbf{k}/\mathbf{k}^2)\, (d^3\mathbf{k}'/\mathbf{k}'^2)\, d^3\mathbf{r} \approx - \tfrac{3}{4}\, \zeta^2\, (m/M)^2\, \Omega^4\, Z\alpha\, \alpha_s\, (1/2\pi)^3 \int\int e^{-i\mathbf{k}'.\mathbf{r}}\, \{\int e^{i[\mathbf{k}' - \mathbf{k}].\mathbf{r}}\, [(\Upsilon^{\mu} q_{\mu} + m_q)(\Upsilon^{\mu} q'_{\mu} + m_q)/(\mathbf{q}^2 - m_q^2)\, (\mathbf{q}'^2 - m_q^2)]\, d^3\mathbf{r}\}\, (d^3\mathbf{k}/\mathbf{k}^2)\, (d^3\mathbf{k}'/\mathbf{k}'^2), \quad (2)$$

on account that $\mathbf{k} = \mathbf{k}' + \mathbf{q}' - \mathbf{q}$. To proceed further let us assign an ordinal number, T, to account for the type of fundamental fermion either a lepton, $T = T_L = 1$, or a quark[2], $T = T_Q = 2$. Moreover, let us make the ansatz $\zeta = - Y_{wL} \sum (-1)^T\, T\, I_3$, where $Y_{wL}$ and $I_3 = I_{3L}$ or $I_{3Q}$ denote respectively the weak hypercharge of the test lepton and the third component of the weak isospin of the lepton or quark. More generally, $\zeta = - Y_{wtest} \sum (-1)^T\, T\, I_3$, where $Y_{wtest}$ denotes the weak hypercharge of the test particle that is interacting with the nucleus. Therefore, since $Y_w = 0$ for the mesons (quark-antiquark pairs), it follows that the extra-potential $V_L(r)$ will reduce to zero for the pionic or kaonic atoms [28] and then provide no contribution to

[2] Apart from gravity and electromagnetism, such a particle is sensitive to T other fundamental interactions. Moreover, the electric charge ratio [∑ electric charge of "upper quarks" (quarks with $I_3 = ½$)]/[∑ electric charge of the overall quarks] = [⅔ e + ⅔ e + ⅔ e]/[⅔ e + ⅔ e + ⅔ e – ⅓ e – ⅓ e – ⅓ e] = 2 = $T_Q$ and likewise [∑ electric charge of "upper leptons" (leptons with $I_3 = -½$)]/[∑ electric charge of all leptons] = [– e – e – e]/[– e – e – e + 0 + 0 + 0] = 1 = $T_L$.

these exotic mesonic atoms or pionium and kaonium (not observed yet). By restoring $\hbar$ and c and setting[3] $\Lambda = (2\alpha_s)^{1/2}\,[(\tfrac{1}{2}\hbar\Omega)^2/m_q c^2]$, the above relations yield

$V_C(r) = -\,(\hbar c\, Z\alpha/r)$, (3)

$V_L(r) \approx 6\hbar c\, Z\alpha\, \zeta^2\, \Lambda^2\, (m/M)^2\, [-\,(1/2\pi)^3 \int e^{-i\mathbf{k}.\mathbf{r}}\, (d^3\mathbf{k}/\mathbf{k}^4)] = \hbar c\, Z\alpha\, [\sum (-1)^T\, T\, I_3]^2\, \Lambda^2\, (m/M)^2\, r$ (4)

since $\mathbf{q}^2 \ll m_q^2$ and $\mathbf{q}'^2 \ll m_q^2$ in the low energy interactions approximation so that $\int e^{i(\mathbf{k}' - \mathbf{k}).\mathbf{r}}\, [(\Upsilon^\mu q_\mu + m_q)(\Upsilon^\mu q'_\mu + m_q)/(\mathbf{q}^2 - m_q^2)(\mathbf{q}'^2 - m_q^2)]\, d^3\mathbf{r} \approx (1/m_q^2)\, \delta(\mathbf{k}' - \mathbf{k})$ and in addition $(1/2\pi)^3 \int e^{-i\mathbf{k}.\mathbf{r}}\, d^3\mathbf{k}/\mathbf{k}^2 = 1/r$ and $(1/2\pi)^3 \int e^{-i\mathbf{k}.\mathbf{r}}\, d^3\mathbf{k}/\mathbf{k}^4 = -\,\tfrac{1}{6}\, r$.

Now, as one knows, $Y_{wL} = -1$ and $I_{3L} = -\,\tfrac{1}{2}$ for the left-handed charged lepton and $\sum I_{3Q} = \tfrac{1}{2}(N_u - N_d)$ for the left-handed up and down quarks, where $N_u$ and $N_d$ denote respectively the number of up quark and the number of down quark. One finds that $\zeta = \sum (-1)^T\, T\, I_3 = T_Q \sum I_{3Q} - T_L \sum I_{3L} = 2 \sum I_{3Q} - \sum I_{3L} = \tfrac{1}{2} + N_u - N_d$. Thus, the overall static potential, V, reads

$V(r) = -\,(\hbar c\, Z\alpha/r) + \hbar c\, Z\alpha\, (\tfrac{1}{2} + N_u - N_d)^2\, (m/M)^2\, (\Lambda/\hbar c)^2\, r$

$\quad = -\,(\hbar c\, Z\alpha/r) + Z\alpha\, (\tfrac{1}{2} + N_u - N_d)^2\, (\Lambda/Mc^2)^2\, mc^2\, (r/\lambda\!\!\!^{-})$. (5)

## III. Proton and deuteron charge radii from muonic hydrogen and muonic deuterium

Starting from the effective potential (5), one derives the force term, $F_r(r) = -\,\partial V(r)/\partial r$, which reads

$F_r(r) = -\,\partial V(r)/\partial r = -\,Z\alpha\, (\hbar c/r^2)\, [1 + (\tfrac{1}{2} + N_u - N_d)^2\, (\Lambda/Mc^2)^2\, (r/\lambda\!\!\!^{-})^2] = -\,Z\alpha\, \hbar c/r_{eff}^2$, (6)

where we have set

$r_{eff} = r/[1 + (\tfrac{1}{2} + N_u - N_d)^2\, (\Lambda/Mc^2)^2\, (r/\lambda\!\!\!^{-})^2]^{1/2}$. (7)

In the first order approximation, the latter quantity may be set equal to the charge radius (see appendix A), $r_E$. Thus, $r_E \approx r_{eff}$ should always be found less than r. Hence, one finds for the ordinary hydrogen atom,

$r_E = r_p/[1 + (\tfrac{1}{2} + N_u - N_d)^2\, (\Lambda/Mc^2)^2\, (r_p/\lambda\!\!\!^{-}_e)^2]^{1/2}$, (8)

and for the muonic hydrogen atom[4],

---

[3] Let us emphasize that $\Lambda = (2\alpha_s)^{1/2}\,[(\tfrac{1}{2}\hbar\Omega)^2/m_{u,d}c^2] \sim 2\Lambda_{QCD} \sim 400$ MeV and $\hbar\Omega \sim \tfrac{1}{2}\,\Lambda_{QCD} \sim 100$ MeV, where $\Lambda_{QCD} \sim 0.2$ GeV denotes the QCD infrared cutoff and $m_{u,d} = (m_u + m_d)/2 \sim 3.5$ MeV/$c^2$ is the mean of the current u and d quarks masses.

[4] $m_e$ = electron mass = 0.510 998 910(13) MeV/$c^2$ ; $m_\mu$ = muon mass = 105.6583715(35) MeV/$c^2$ ; $m_p$ = proton mass = 938.272013(23) MeV/$c^2$ ; $m_D$ = deuteron mass = 1875.612793(47) MeV/$c^2$ ; $m_T$ = triton mass =

$r_E = r_p/[1 + (½ + N_u – N_d)^2 (\Lambda/Mc^2)^2 (r_p/\bar{\lambda}_\mu)^2]^{1/2}$. (9)

For ${}_1H^1$, one has $N_u = 2$, $N_d = 1$ and $½ + N_u – N_d = 3/2$. Hence $(\Lambda/\hbar c)^2 = 3.862\times10^{30}$ m$^{-2}$.

For ${}_1H^2$, one has $N_u = 3$ and $N_d = 3$ so that $½ + N_u – N_d = ½$. Hence $(\Lambda/\hbar c)^2 = 4.350\times10^{30}$ m$^{-2}$.

So, we derive for both measurements $(\Lambda/\hbar c)^2 = (4.106 \pm 0.244)\times10^{30}$ m$^{-2}$ and hence $\Lambda = (0.400 \pm 0.012)$ GeV, which implies a relative uncertainty of only 3 %.

Since the proton radius is equal to $r_p = 0.8768(69)$ fm [1], one finds $(r_E)_p \approx r_p$ for the ordinary hydrogen atom and $(r_E)_p = 0.84184(67)$ fm for the muonic hydrogen atom, with a relative uncertainty of 0.3%. For the deuterium atom, one has A = 2 and $r_D = 2.1424(21)$ fm, thus the above relation yields for the muonic deuterium atom $(r_E)_D = 2.12562(78)$ fm which matches quite well with the experimental result [29]. For the tritium atom, one has A = 3 and $r_T = 1.7591 \pm 0.0363$ fm, thus for the muonic tritium atom we predict from relation (11), $(r_E)_T = 1.7562$ fm which lies within the range of the present experimental value of $r_T$ up to uncertainties. Likewise, for ${}_6C^{12}$, one has $N_u = N_d = 18$ so that $½ + N_u – N_d = ½$ and $r_E$(C12) = $r_{eff}$(ordinary C12) = 2.475 fm and $r_E$(μC12) = $r_{eff}$(μC12) = 2.474 fm in good agreement with the experimental data [30,31,32,33,34].

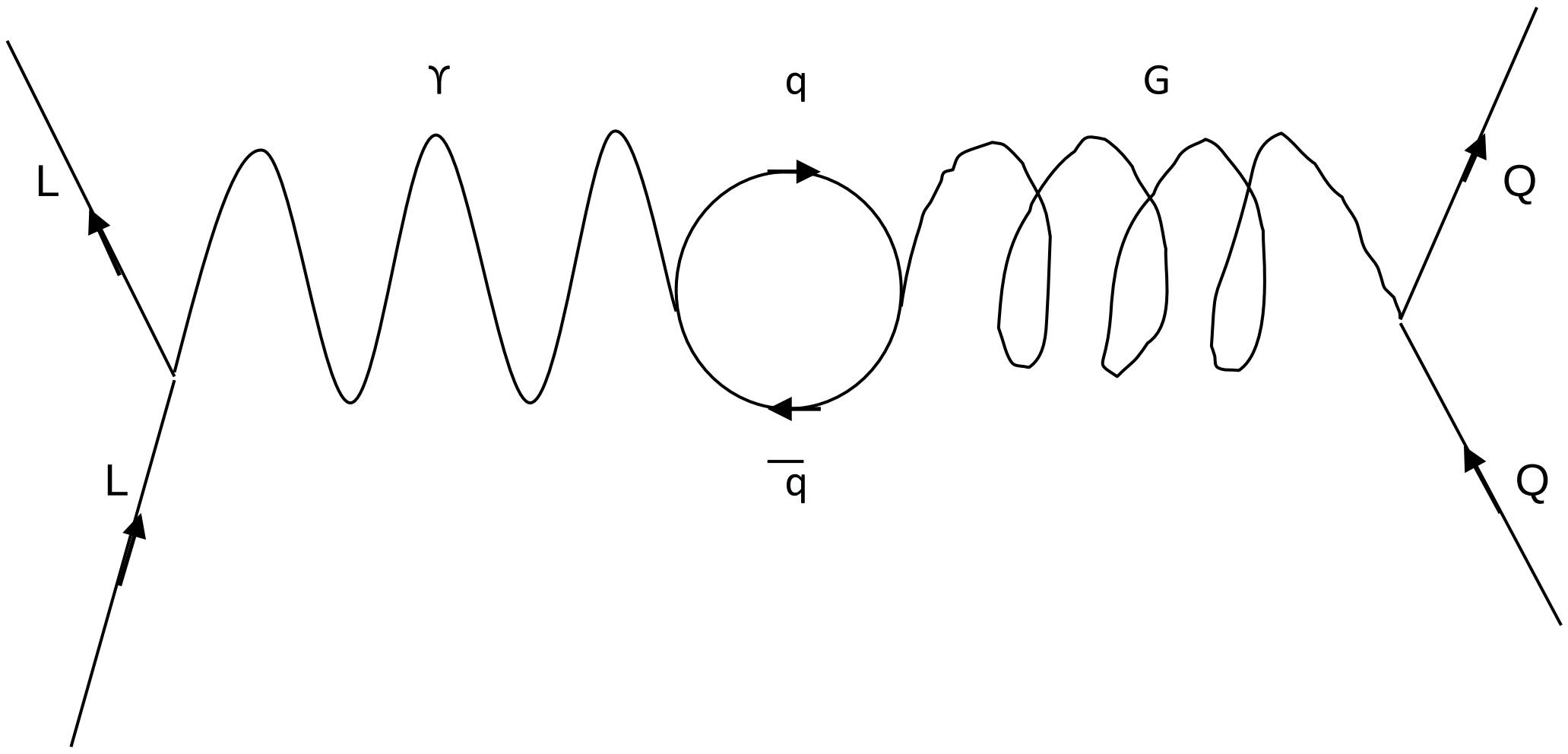

FIG. 1 : A second order diagram involving a loop (vacuum polarisation effect and screening correction), containing a virtual quark-antiquark pair with both QED and QCD coupling vertices respectively through lepton-photon and quark-gluon couplings, an incoming and outgoing lepton, L, as well as an incoming and outgoing quark, Q, which denotes either a down or up quark of a nucleon of a light nucleus. We argue that the effective potential associated to such a mixing interaction reads $V(r) = – (\hbar c Z\alpha/r) + \hbar c Z\alpha\, \alpha_s\, \zeta^2 (m/M)^2 (r/\bar{\lambda}_0)$, by setting $\bar{\lambda}_0 = (\hbar/M_0c)$ and $M_0c^2 = \Lambda\, \alpha_s^{-1/2}$, where $\alpha_s$ denotes the strong coupling constant at the energy scale $4.5 \sqrt{2}\, M_0c^2$ ; $\alpha_s = 0.16$ for $M_0 = 1$ GeV/c$^2$.

---

2808.921112(17) MeV/c$^2$ ; $m_\pi$ = neutral pion mass = (134,9766 ± 0,0006) MeV/c$^2$ ; $\bar{\lambda}_e = 3.8615926459(53)\times10^{-13}$ m ; $\bar{\lambda}_\mu = 1.867594294(47)\times10^{-15}$ m.

IV. The magnetic radius of ordinary and muonic hydrogen

The Zemach radius, $r_Z = \int d^3\mathbf{r}\, r \int d^3\mathbf{r}'\, \rho_E(\mathbf{r} - \mathbf{r}')\, \rho_M(\mathbf{r}')$ = 1.082(37) fm, as derived from the 2S hyperfine splitting $\Delta E_{HFS}$ = 22.9763(15) – 0.1621(10) $r_Z$ + 0.0080(26) meV [2] of both transitions $2S_{1/2}^{F=1}$ to $2P_{3/2}^{F=2}$ and $2S_{1/2}^{F=0}$ to $2P_{3/2}^{F=1}$, and the magnetic rms radius, $r_M$ = 0.87(6) fm, of the proton derived from the spectroscopy of muonic hydrogen both remain compatible with that obtained from either electron-proton scattering or normal hydrogen spectroscopy. However, the effective potential V(r) implies a spin-orbit correction. Indeed, the magnetic field acting on the test charged lepton reads

$$\mathbf{B} = (1/mc^2)\,(1/r)\,[\partial V(r)/\partial r]\,\mathbf{L} = -Z\alpha\,(\bar{\lambda}/r^3)\,[1 + (½ + N_u - N_d)^2\,(\Lambda/Mc^2)^2\,(r/\bar{\lambda})^2]\,\mathbf{L}$$

$$= -Z\alpha\,(\bar{\lambda}/r_{eff}^3)\,\mathbf{L}, \quad (10)$$

where we have set

$$r_{eff} = r/[1 + (½ + N_u - N_d)^2\,(\Lambda/Mc^2)^2\,(r/\bar{\lambda})^2]^{1/3}. \quad (11)$$

In the first order approximation, the latter quantity may be set equal to the magnetic radius, $r_M$. Hence, one finds for the ordinary hydrogen atom,

$$r_M = r_p/[1 + (½ + N_u - N_d)^2\,(\Lambda/Mc^2)^2\,(r_p/\bar{\lambda}_e)^2]^{1/3}, \quad (12)$$

and for the muonic hydrogen atom

$$r_M = r_p/[1 + (½ + N_u - N_d)^2\,(\Lambda/Mc^2)^2\,(r_p/\bar{\lambda}_\mu)^2]^{1/3}. \quad (13)$$

Thus, $r_M$ should be found slightly less than $r_p$, but for precise measurements. Otherwise, $(r_M)_p \approx r_p$ for the ordinary hydrogen atom and $(r_M)_p$ = 0.8519(22) fm for the muonic hydrogen atom which matches quite well with the experimental result $r_M$ = 0.87(6) fm [2] and most of the previous experimental data [7, 29, 35] but see also [36,37]. The rms magnetic radius of the ordinary deuteron and triton too match quite well with the corresponding muonic atoms counterparts which are predicted to be respectively $(r_M)_D$ = 2.1318(21) fm and $(r_M)_T$ = 1.7565 ± 0.0362 fm.

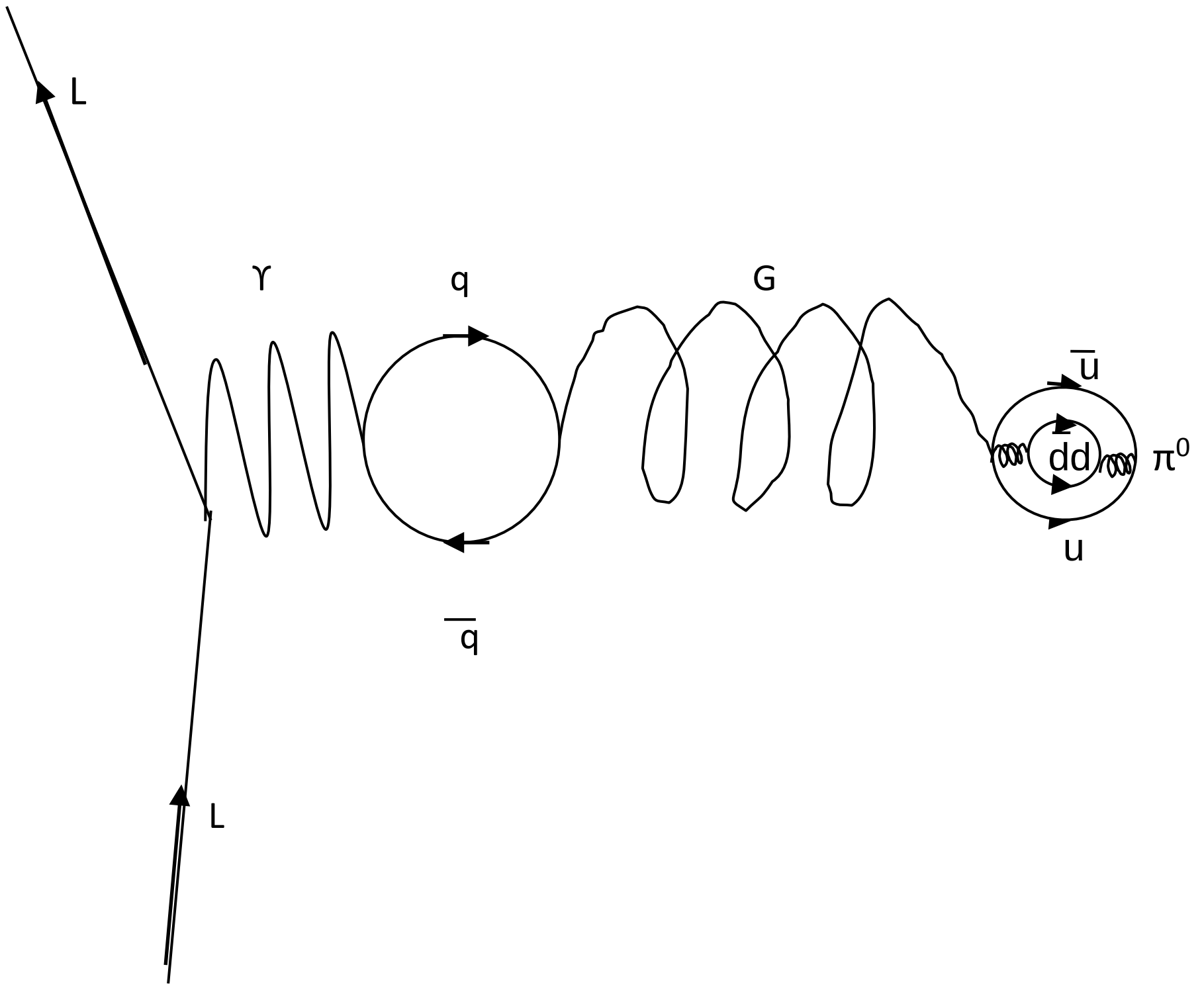

FIG. 2 : Almost the same second order diagram as in FIG. 1, where the incoming and outgoing quark is replaced by a pair of either an up-anti up or down-anti down quarks of a virtual neutral pion, $\pi^0$.

V. The anomalous magnetic moment

Finally, the same approach may help to solve the problem of the anomalous magnetic moment of the muon too. From the relations $F_r(r) = -Z\alpha\,(\hbar c/r^2)\ [1 + (½ + N_{quark} - N_{antiquark})^2 (\Lambda/Mc^2)^2 (r/ƛ)^2]$ or $\mathbf{B} = -Z\alpha\,(ƛ/r^3)\ [1 + (½ + N_{quark} - N_{antiquark})^2 (\Lambda/Mc^2)^2 (r/ƛ)^2]\,\mathbf{L}$ (see, FIG. 2), one may define an effective fine structure constant, $\alpha_{eff}$, such that $F_r(r) = -Z\alpha_{eff}\,(\hbar c/r^2)$ or $\mathbf{B} = -Z\alpha_{eff}\,(ƛ/r^3)\,\mathbf{L}$. In both cases, one obtains

$\alpha_{eff} = \alpha\,[1 + (\Lambda/2Mc^2)^2 (r/ƛ)^2]$, (14)

since $N_{antiquark} = N_{quark}$. Thus, it follows an effective anomalous magnetic moment for the charged lepton under consideration [38],

$a_{eff} = \alpha_{eff}/2\pi = (\alpha/2\pi) + (\alpha/2\pi)\,(\Lambda/2Mc^2)^2 (r/ƛ)^2 = a + \delta a$, (15)

where $a = \alpha/2\pi$ and

$\delta a = (\alpha/2\pi)\,(\Lambda/2Mc^2)^2 (r/ƛ)^2$. (16)

Here, the virtual pion $\pi^0$ replace the nucleus so that $M \approx m_\pi = (134{,}9766 \pm 0{,}0006)$ MeV/$c^2$. Similarly, the radius $r \approx (\hbar/\mu c) \approx 2.164\times10^{-18}$ m, where we have set $\mu = \sum[m_{quark}m_{antiquark}/(m_{quark} + m_{antiquark})] = ½ \sum m_{quark} = ½\,(m_u + m_d + m_s + m_c + m_b + m_t) \approx 91$

GeV/$c^2$ ≈ $M_Z$ [39], provides an estimate of the distance between the muon and the virtual pion $\pi^0$. Hence, in the case of the muon, $\delta a_\mu \approx (\alpha/8\pi)\,(\Lambda/m_\pi c^2)^2\,(m_\mu/M_Z)^2 \approx (342 \pm 20) \times 10^{-11}$, in consistency with the experimental deviation $\delta a_\mu = (a_\mu)_{exp} - (a_\mu)_{the} = (302 \pm 87) \times 10^{-11}$.

## VI. Conclusion

We have shown that an effective potential which includes a linear extra-potential may solve both the proton and deuteron size puzzle as well as the anomalous magnetic moment of the muon. This is achieved while still preserving the lepton universality and being consistent with both the rms magnetic radius of the proton and deuteron, the rms charge radius of the ordinary or muonic C12 atom and the pionic and kaonic X-rays spectroscopy as well. The relations derived in this paper are predictive and in accordance with the experimental data.

## Appendix A

In the first order approximation, the charge density of the proton (uud), may read

$\rho_E(\mathbf{r}) = ⅔\, e\, \delta(\mathbf{r} - \mathbf{r}_{u1}) + ⅔\, e\, \delta(\mathbf{r} - \mathbf{r}_{u2}) - ⅓\, e\, \delta(\mathbf{r} - \mathbf{r}_d)$,

Throughout, $\mathbf{r}_Q$ denotes the position vector between the center of mass of the nucleus and the position of a quark, Q, within the given nucleus. Hence,

$<r_E^2> = \int r^2\, (\rho_E(\mathbf{r})/e)\, d^3\mathbf{r}$

$= \int r^2\, [⅔\, \delta(\mathbf{r} - \mathbf{r}_{u1}) + ⅔\, \delta(\mathbf{r} - \mathbf{r}_{u2}) - ⅓\, \delta(\mathbf{r} - \mathbf{r}_d)]\, d^3\mathbf{r} = ⅔\, r_{u1}^2 + ⅔\, r_{u2}^2 - ⅓\, r_d^2$.

Now, $\mathbf{r}_Q = \mathbf{r}_{eff} + \delta\mathbf{r}_Q$, so that $<r_Q^2> = r_{eff}^2 + \delta\mathbf{r}_Q^2 \approx r_{eff}^2$, since $(\delta\mathbf{r}_Q/r_{eff})^2 \ll 1$ and $<\mathbf{r}_{eff} \,.\, \delta\mathbf{r}_Q> = 0$.

Therefore, $<r_E^2> \approx ⅔\, r_{eff}^2 + ⅔\, r_{eff}^2 - ⅓\, r_{eff}^2 = r_{eff}^2$.

In the first order approximation, the charge density of the deuteron (uud + udd), may read

$\rho_E(\mathbf{r}) = ⅔\, e\, \delta(\mathbf{r} - \mathbf{r}_{u1}) + ⅔\, e\, \delta(\mathbf{r} - \mathbf{r}_{u2}) - ⅓\, e\, \delta(\mathbf{r} - \mathbf{r}_{d3}) - ⅓\, e\, \delta(\mathbf{r} - \mathbf{r}_{d1}) - ⅓\, e\, \delta(\mathbf{r} - \mathbf{r}_{d2}) + ⅔\, e\, \delta(\mathbf{r} - \mathbf{r}_{u3})$.

Hence,

$<r_E^2> = \int r^2\, (\rho_E(\mathbf{r})/e)\, d^3\mathbf{r}$

$= \int r^2\, [⅔\, \delta(\mathbf{r} - \mathbf{r}_{u1}) + ⅔\, \delta(\mathbf{r} - \mathbf{r}_{u2}) - ⅓\, \delta(\mathbf{r} - \mathbf{r}_{d3}) - ⅓\, e\, \delta(\mathbf{r} - \mathbf{r}_{d1}) - ⅓\, e\, \delta(\mathbf{r} - \mathbf{r}_{d2}) + ⅔\, e\, \delta(\mathbf{r} - \mathbf{r}_{u3})]\, d^3\mathbf{r}$

$= ⅔\, r_{u1}^2 + ⅔\, r_{u2}^2 - ⅓\, r_{d3}^2 - ⅓\, r_{d1}^2 - ⅓\, r_{d2}^2 + ⅔\, r_{u3}^2$

$\approx ⅔\, r_{eff}^2 + ⅔\, r_{eff}^2 - ⅓\, r_{eff}^2 - ⅓\, r_{eff}^2 - ⅓\, r_{eff}^2 + ⅔\, r_{eff}^2 = r_{eff}^2$.

In the first order approximation, the charge density of the helion (2uud + 2udd), may read

$\rho_E(\mathbf{r})$ = ⅔ e $\delta(\mathbf{r} - \mathbf{r}_{u1})$ + ⅔ e $\delta(\mathbf{r} - \mathbf{r}_{u2})$ – ⅓ e $\delta(\mathbf{r} - \mathbf{r}_{d3})$ – ⅓ e $\delta(\mathbf{r} - \mathbf{r}_{d1})$ – ⅓ e $\delta(\mathbf{r} - \mathbf{r}_{d2})$

+ ⅔ e $\delta(\mathbf{r} - \mathbf{r}_{u3})$ + ⅔ e $\delta(\mathbf{r} - \mathbf{r}_{u'1})$ + ⅔ e $\delta(\mathbf{r} - \mathbf{r}_{u'2})$ – ⅓ e $\delta(\mathbf{r} - \mathbf{r}_{d'3})$

– ⅓ e $\delta(\mathbf{r} - \mathbf{r}_{d'1})$ – ⅓ e $\delta(\mathbf{r} - \mathbf{r}_{d'2})$ + ⅔ e $\delta(\mathbf{r} - \mathbf{r}_{u'3})$.

Hence,

$<r_E^2> = \int r^2 (\rho_E(\mathbf{r})/2e)\, d^3\mathbf{r}$

= ½ $\int r^2$ [⅔ $\delta(\mathbf{r} - \mathbf{r}_{u1})$ + ⅔ $\delta(\mathbf{r} - \mathbf{r}_{u2})$ – ⅓ $\delta(\mathbf{r} - \mathbf{r}_{d3})$ – ⅓ e $\delta(\mathbf{r} - \mathbf{r}_{d1})$ – ⅓ e $\delta(\mathbf{r} - \mathbf{r}_{d2})$

+ ⅔ e $\delta(\mathbf{r} - \mathbf{r}_{u3})$ + ⅔ e $\delta(\mathbf{r} - \mathbf{r}_{u'1})$ + ⅔ e $\delta(\mathbf{r} - \mathbf{r}_{u'2})$ – ⅓ e $\delta(\mathbf{r} - \mathbf{r}_{d'3})$ – ⅓ e $\delta(\mathbf{r} - \mathbf{r}_{d'1})$

– ⅓ e $\delta(\mathbf{r} - \mathbf{r}_{d'2})$ + ⅔ e $\delta(\mathbf{r} - \mathbf{r}_{u'3})$] $d^3\mathbf{r}$ = ½ (⅔ $r_{u1}^2$ + ⅔ $r_{u2}^2$ – ⅓ $r_{d3}^2$ – ⅓ $r_{d1}^2$ – ⅓ $r_{d2}^2$

+ ⅔ $r_{u3}^2$ + ⅔ $r_{u'1}^2$ + ⅔ $r_{u'2}^2$ – ⅓ $r_{d'3}^2$ – ⅓ $r_{d'1}^2$ – ⅓ $r_{d'2}^2$ + ⅔ $r_{u'3}^2$)

≈ ⅔ $r_{eff}^2$ + ⅔ $r_{eff}^2$ – ⅓ $r_{eff}^2$ – ⅓ $r_{eff}^2$ – ⅓ $r_{eff}^2$ + ⅔ $r_{eff}^2$ = $r_{eff}^2$.

References

[1] Pohl R. *et al.*, *Nature*, **466** (2010) 213.

[2] Antognini A. *et al.*, *J. Phys. :Conf. Ser.*, **312** (2011) 032002.

[3] Antognini A. *et al.*, *Science*, **339** (2013) 417.

[4] Mohr P. J. *et al.*, Reviews of Modern Physics, **80**, (2008) 633.

[5] Mohr P. J. *et al.*, Reviews of Modern Physics, **84**, (2012).

[6] Mohr P. J. *et al.*, Reviews of Modern Physics, **88**, (2016).

[7] Zhan X. *et al.*, *Phys. Lett. B* **705** (2011) 59.

[8] Onofrio R., *EPL*, **104** (2013) 20002.

[9] M. Haghighat and M. Khorsandi, arXiv:1410.0836v2.

[10] Yu-Sheng Liu and G. A. Miller, Phys. Rev. C **92** (2015) 035209 ; Yu-Sheng Liu, D. McKeen and G. A. Miller, Phys. Rev. Lett. 117, 101801 (2016).

[11] Dahia, F. and Lemos, A.S., Eur. Phys. J. C **76** (2016) 435.

[12] L.-B. Wang and W.-T. Ni, Mod. Phys. Lett. A 28 (2013) 1350094.

[13] M. Eides, (2014), Proton Radius Puzzle: New Physics?, http://hep.phys.spbu.ru/conf/novozhilov90/materials.html

[14] V. Barger et al., Phys.Rev.Lett. **106** (2011) 153001.

[15] F. Jegerlehner and A. Nyffeler, (2009), Physics Report **477** (2009) 1.

[16] T. Kinoshita and W. J. Marciano, in Quantum Electrodynamics, edited by T. Kinoshita (World Scientific, Singapore, 1990), pp. 419–478 ; A. Czarnecki and W. J. Marciano, Phys. Rev. D **64** (2001) 013014 ; D. Stöckinger, J. Phys. G **34** (2007) R45 ; T. P. Gorringe and D.W. Hertzog, arXiv:1506.01465 [hep-ex].

[17] M. I. Eides *et al.*, Phys. Rept. **342** (2001) 63.

[18] A. Antognini, Annals Phys. **331** (2013) 127.

[19] F. Jegerlehner, (2012), The Anomalous Magnetic Moment of the Muon, Monograph, Springer

[20] G.W. Bennett *et al.*, Phys. Rev. D **73** (2006) 072003 ; G.W. Bennett et al., Phys. Rev. Lett. **92** (2004) 161802 ; G.W. Bennett *et al.*, Phys. Rev. Lett. **89** (2002) 101804 ; G.W. Bennett *et al.*, Phys. Rev. Lett. **89** (2002) 129903 (Erratum) ; Passera M., *Phys. Rev. D*, **75** (2007) 013002.

[21] K. Hagiwara *et al.*, Phys. Lett. B **649**, (2007) 173.

[22] M. Davier *et al.*, Eur. Phys. J. C **71** (2011) 1515 ; A. Hoecker & W. J. Marciano, (2013), The muon anomalous magnetic moment, PDG.

[23] Kinoshita T. and Nio M., Phys. Rev. D, **70** (2004) 113001.

[24] Passera M., *Phys. Rev. D*, **75** (2007) 013002.

[25] Kinoshita T. *et al.*, *Phys. Rev. D*, **31** (1985) 2108.

[26] Czarnecki A., Marciano W. J. and Vainshtein A., *Phys. Rev. D*, **67** (2003) 073006.

[27] Serenone W. M. *et al.*, (2016), Heavy-Quarkonium Potential from the Lattice Gluon Propagator, Journal of Physics: Conference Series 706, 052038 ; arXiv:1505.06720.

[28] Bazzi M. et *al.*, Phys. Lett. B **704**, (2011) 113 ; Shi H., EPJ Web Conf. **126**, (2016), arXiv:1601.02236 ; Bazzi M. et *al.*, Nucl. Phys. A **954**, (2016) 7 ; Curceanu C. et *al.*, Eur. Phys. J. A **31** (2007) 537 ; Indelicato P. *et al.*, Phys. Rev. Lett. **91** (2003) 240801 ; Gotta D., Europhysics News **43** (2012) 28.

[29] Pohl R. *et al.*, Science, **353** (2016) 669.

[30] L.S. Cardman *et al.*, Phys. Lett. B **91**, (1980) 203.

[31] Offermann E.A.J. M. *et al.*, Phys. Rev. C **44**, (1991) 1096.

[32] Ruckstuhl W. *et al.*, Nucl. Phys. A **430** (1984) 685.

[33] Schaller L. A. *et al.*, Nucl. Phys. A **379** (1982) 523.

[34] Sanford T. *et al.*, Phys. Rev. C **8**, (1973) 896.

[35] D. Borisyuk, (2010), Nucl. Phys. A **843**, 59.

[36] I. T. Lorenz, H.-W. Hammer and U. Meißner, Eur. Phys. J. A **48** (2012) 151.
[37] J. C. Bernauer *et al.*, Phys. Rev. C **90** (2014) 015206.
[38] J. S. Schwinger, Phys. Rev. **73** (1948) 416.
[39] C. Patrignani *et al.* (Particle Data Group), *Chin. Phys. C* **40** (2016) 100001.